\documentclass{article}
\usepackage{graphicx}
\graphicspath{%
    {converted_graphics/}
    {/}
}

\begin{document}

\begin{center}
{\LARGE \ Chemical Influences on Ice Crystal Growth from Vapor}\vskip6pt

{\Large Kenneth Libbrecht and Robert Bell}\vskip4pt

{\large Department of Physics, California Institute of Technology}\vskip-1pt

{\large Pasadena, California 91125}\vskip-1pt

{\large address correspondence to: kgl@caltech.edu}\vskip-1pt

\vskip18pt

\hrule\vskip1pt \hrule\vskip14pt
\end{center}

\noindent \textbf{Abstract. We present an investigation of chemical
influences on the growth of ice crystals from water vapor. In one set of
experiments, we grew ice crystals in a vapor diffusion chamber, observing
crystal morphologies at temperatures from 0 C to -25 C in different
background gases and with a variety of gaseous chemical additives. In a
second set of experiments, we measured ice crystal growth rates at -5 C and
-15 C in a free-fall flow chamber, using normal laboratory air and
ultra-clean nitrogen gas, both with and without chemical additives.
Conclusions from these experiments include:\ 1) In nitrogen gas at a
pressure of one atmosphere, no tested chemical additives at concentrations
below 10 ppm produced any observable changes in ice crystal growth
morphologies; 2) Growth in ultra-clean nitrogen gas was not significantly
different from growth in ordinary laboratory air; 3) Chemical additives
affected plate-like dendritic growth at -15 C more readily than growth at
higher temperatures; 4) Chemical additives tended to promote the growth of
columnar crystals over plate-like crystals; 5) Ice growth in air, nitrogen,
helium, argon, hydrogen, and methane gases at a pressure of one atmosphere
yielded essentially the same temperature-dependent crystal morphologies. }

\vskip4pt \noindent \textit{[The figures in this paper have been reduced in
size to facilitate rapid downloading. The paper is available with higher
resolution figures at http://www.its.caltech.edu/\symbol{126}%
atomic/publist/kglpub.htm, or by contacting the authors.]}

\section{Introduction}

The formation of snow crystals -- ice crystals that grow from water vapor in
the atmosphere -- is a remarkably complex process that produces a rich
phenomenology of growth structures and patterns. Although researchers have
documented snow crystal morphologies for centuries, and have studied ice
growth under controlled conditions for over 75 years, our basic
understanding of ice crystal growth dynamics remains poor \cite%
{libbrechtreview}. For example, the observed change in ice growth rates and
morphologies with temperature has yet to be explained at even a qualitative
level.

One hindrance to the study of ice crystal growth dynamics is the influence
of unwanted chemical impurities that are always present at some level in
every experimental apparatus. Vapor impurities in the background gas
surrounding a growing ice crystal can attach to the crystal surface and
affect its growth in unanticipated ways. If the impurity concentrations are
not known and are not controlled, then ice growth measurements may yield
unreliable results.

Although much work has been done on the interaction of trace gases in the
atmosphere with ice surfaces \cite{abbatt}, relatively little systematic
research has investigated how surface chemistry affects ice crystal growth.
Nevertheless, it has been known for decades that chemical additives can
dramatically change ice growth rates and morphologies. Vonnegut \cite%
{vonnegut, mason, hallettandmason} found that while plates are the normal
growth morphology at -20 C in air, the addition of 10 ppm of butyl alcohol
yielded columnar growth instead. Schaefer \cite{schaefer, mason} further
observed that vapors of ketones, fatty acids, silicones, aldehydes, and
alcohols could all change ice growth morphologies to varying degrees.
Nakaya, Hauajima, and Mugurama \cite{nakaya2} observed that even trace
silicone vapor in air caused columnar crystals to grow at -15C, while the
usual habit at this temperature is plate-like. Hallett and Mason \cite%
{hallettandmason} found that the addition of camphor vapor in air could
yield columnar ice crystals at all temperatures in the range $-40$ C 
$<$ $T$ $<$ $0$ C. These authors also observed that
isobutyl alcohol in air changed ice growth near -15 C from plates to columns
and then back to plates again as the concentration was increased. Anderson,
Sutkoff, and Hallett \cite{anderson} found that Methyl 2-Cyanoacrylate in
air could change the morphology from plate-like dendrites to needles at -15
C. Libbrecht, Crosby, and Swanson \cite{electric} found that acetic acid and
other vapors promoted the c-axis growth of \textquotedblleft
electric\textquotedblright\ needle crystals in air near -5 C. And Knepp,
Renkens and Shepson \cite{knepp} observed various morphological changes
caused by acetic acid vapor in air, even in concentrations as low as 1 ppm.
Given our poor understanding of ice growth without the complicating effects
of chemical additives, it should come as no surprise that there is
essentially no theoretical understanding, even at a qualitative level, of
how chemical additives alter growth rates and change growth morphologies.

The current investigation was undertaken in response to a speculative
hypothesis put forth by Libbrecht \cite{impurities}, postulating that even
the low levels of chemical impurities that are normally present in ordinary
air could profoundly affect ice growth rates, and could explain the
enigmatic snow crystal morphology diagram. This hypothesis predicted that
ice growth rates would increase rather dramatically in extremely clean air,
so we set out to test this prediction. In addition to producing an
especially clean growth environment, we also examined ice growth as various
chemical vapors were added to the background air, in an effort to produce a
more quantitative survey of chemical influences on ice crystal growth.

\section{Diffusion Chamber Experiments}

In one set of experiments we observed ice growth in a water vapor diffusion
chamber operating at a pressure of one atmosphere. We used this chamber to
investigate morphological changes as a function of concentration for a
variety of chemical additives in nitrogen gas, and to examine ice growth in
other pure gases. The diffusion chamber has the advantage that one can
simultaneously observe ice growth over a range of temperatures, in our case
from $-25$ C $<$ $T$ $<$ 0 C. One disadvantage,
however, is that the supersaturation varies with location in the chamber and
is thus is only approximately known.

A schematic diagram of our diffusion chamber is shown in Figure \ref%
{diffchamber}. It is essentially a sealed cylindrical chamber with an inside
height of 36 cm and an inside diameter of 20 cm (total volume 11.3 liters),
with inside surfaces made primarily from acrylic, aluminum, and copper.
During operation the top of the chamber was heated to 45 C while the bottom
was cooled to -38 C, and the entire chamber was thermally insulated from its
surroundings. Water vapor diffusing down from the water reservoir produced a
highly supersaturated environment inside the chamber, and a removable
monofilament nylon string with a diameter of 0.2 mm was placed along the
axis of the chamber to nucleate the growth of ice crystals. A typical run
with this chamber lasted \symbol{126}10 hours, during which the string was
wiped and replaced approximately every 60-90 minutes. After each run the
chamber was opened and baked for 1-2 days to remove residual chemicals.

\begin{figure}[h] 
  \centering
  \includegraphics[bb=0 0 996 1069,width=5in,keepaspectratio]{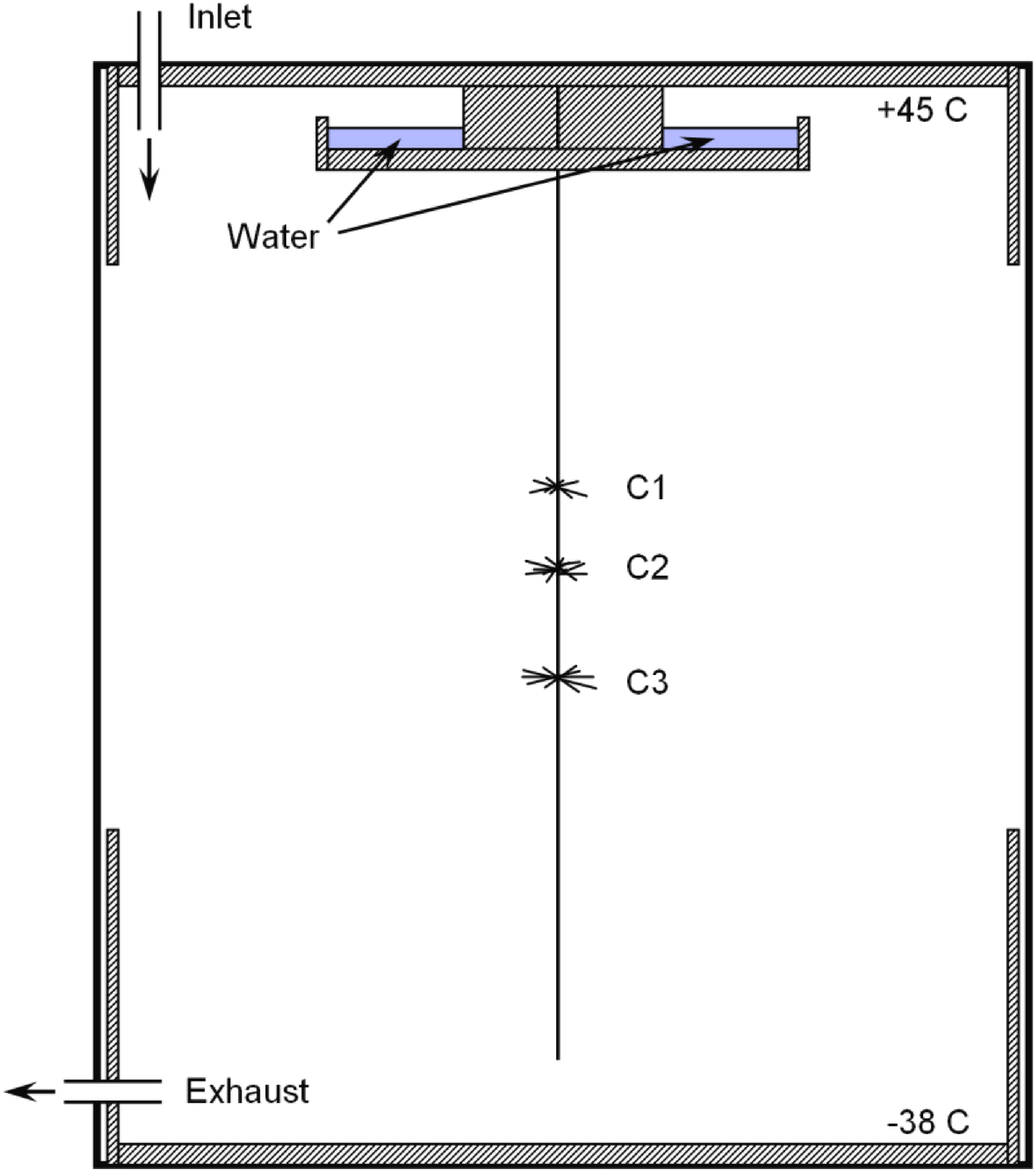}
  \caption{A schematic diagram of the diffusion chamber used in the
present observations, which operated at a pressure of one atmosphere. The
labels C1, C2, and C3 refer to the main clusters of ice crystals growing at
temperatures near -2C, -5C, and -15C, as described in the text. Nitrogen gas
with a variety of vaporous chemical additives was flowed slowly through the
chamber from top to bottom.}
  \label{diffchamber}
\end{figure}

\begin{figure}[h] 
  \centering
  \includegraphics[bb=0 0 877 1035,width=5in,keepaspectratio]{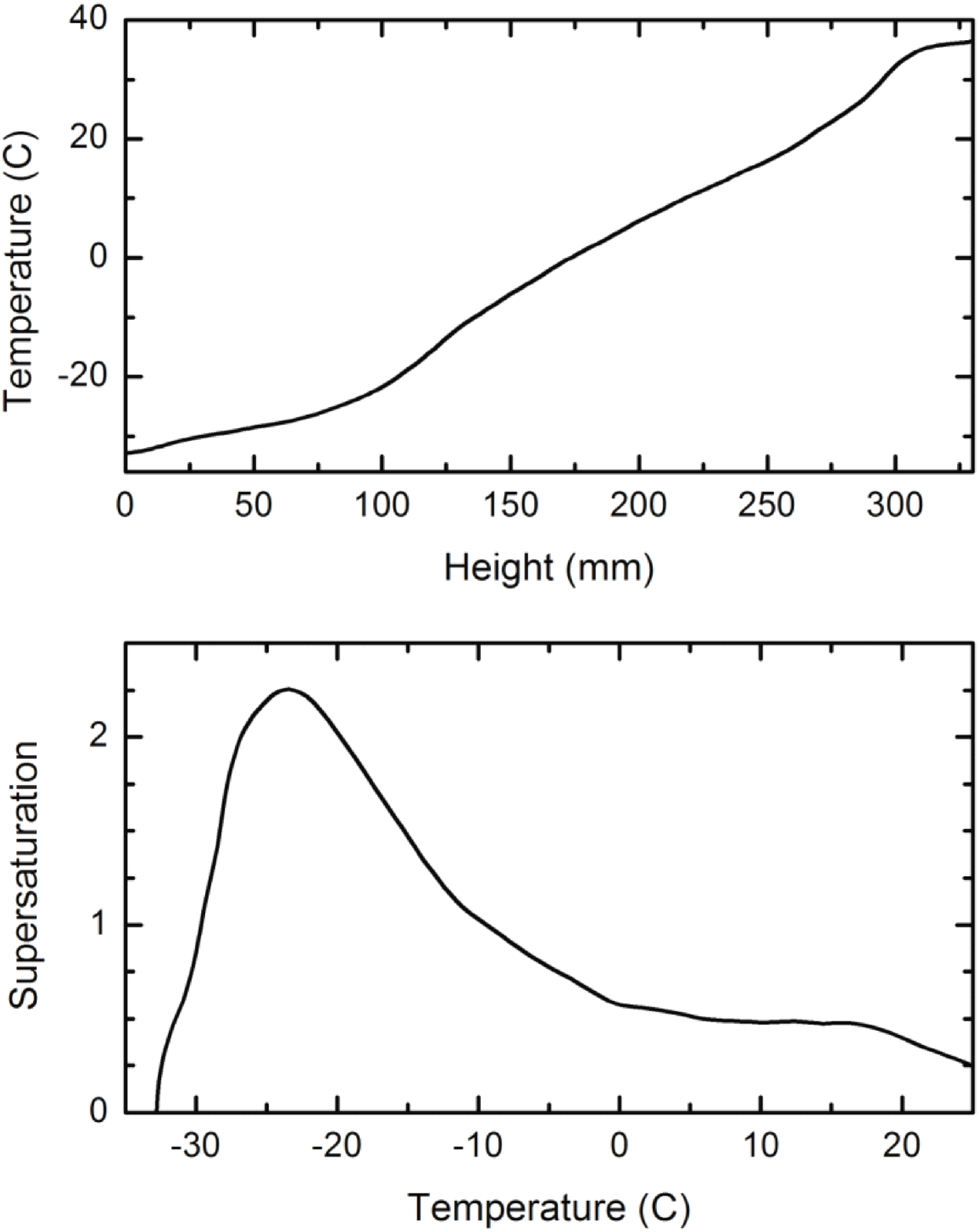}
  \caption{The measured on-axis temperature
profile $T(z)$ inside the diffusion chamber (top), where $z$ is height above
the base of the chamber, and the calculated on-axis supersaturation $\protect%
\sigma _{calc}(T)$ (bottom). As described in the text, the actual
supersaturation inside the chamber was depleted by the presence of growing
crystals, and was thus substantially lower than $\protect\sigma _{calc}(T).$}
  \label{diffchamberprofile}
\end{figure}

The measured on-axis temperature profile $T(z)$ of the diffusion chamber
(where $z$ is height above the base of the chamber) is shown in Figure \ref%
{diffchamberprofile}. Metal plates (shaded in Figure \ref{diffchamber})
reduced condensation near the top of the chamber and produced an especially
steep temperature gradient (and thus high supersaturation) at temperatures
near $T=-20$ C. A calculated on-axis water-vapor supersaturation $\sigma
_{calc}(T)$ in the chamber is also shown in Figure \ref{diffchamberprofile}.
Note that while the temperature profile $T(z)$ was easily measured \textit{%
in situ}, the supersaturation profile $\sigma _{calc}(T)$ shown in Figure %
\ref{diffchamberprofile} was determined by modeling diffusion inside the
chamber, and we expect that the modeled $\sigma _{calc}(T)$ is only a rough
approximation of the true supersaturation $\sigma _{true}(T)$ seen by the
growing crystals.

The largest perturbation to the calculated supersaturation was the presence
of ice crystals growing on the string, which substantially lowered the
on-axis supersaturation. Comparing the growth velocities of needle crystals
with \cite{electric}, we estimate that crystals extending the farthest from
the string experienced a supersaturation $\sigma _{true}(T)\approx \sigma
_{calc}(T)/2$. Crystals growing closer to the string, which were shadowed by
neighboring crystals, experienced substantially lower supersaturations. In
addition to this effect, there were likely weak convection currents in the
chamber that affected the supersaturation profile even in the absence of ice
crystals. Because of this uncertainly in the supersaturation, our diffusion
chamber observations were mainly qualitative in nature, as discussed below.
We were able, however, to crudely examine ice crystal growth as a function
of supersaturation in our experiments by observing morphological differences
between shadowed and less shadowed crystals.

\subsection{Baseline Crystal Morphologies}

\begin{figure}[h] 
  \centering
  \includegraphics[bb=0 0 814 1460,width=3.2in,keepaspectratio]{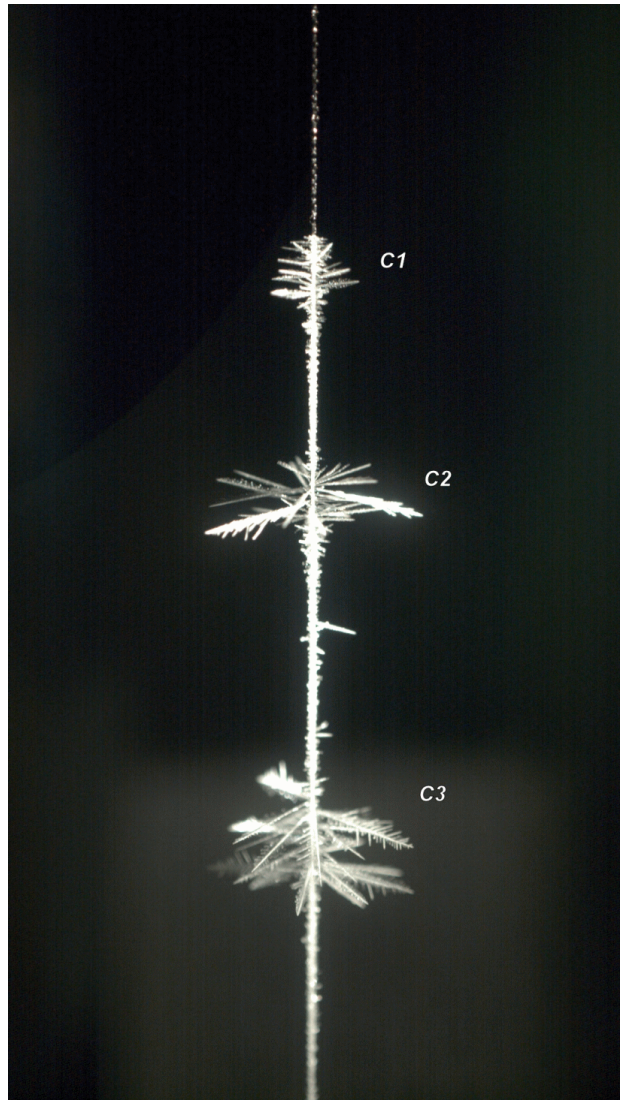}
  \caption{A photograph of ice crystals
growing on a nylon string in the diffusion chamber. Labels show the main
crystal clusters described in the text.}
  \label{string}
\end{figure}

In the absence of any chemical additives, we observed the usual three
dominant clusters of ice crystals growing on the string \cite%
{hallettandmason, mason} in air or nitrogen at one atmosphere, which we
label C1, C2, and C3 in Figures \ref{diffchamber} and \ref{string}. One hour
after placing the string, we observed the following:

C1 -- A small cluster of planer dendritic crystals growing at $T\approx -2$
C. This cluster was typically first visible after 45-60 minutes of growth,
and at one hour the longest crystals were about 2-3 mm in length. The
morphology was mainly plate-like dendritic crystals with sidebranching,
although crystals growing in lower-$\sigma $ regions (shadowed by other
crystals) exhibited a sectored-plate morphology. Condensed water droplets
were observed above C1 on the string, while small sectored plates were seen
immediately below C1.

C2 -- A well defined cluster of needle-like crystals centered at $T\approx
-5 $ C. This cluster was usually visible after 10-15 minutes, and the
vertical width of the cluster was about 2 mm, implying a temperature range
of about $\Delta T\approx 0.7$ C. At one hour the longest crystals were
about 6 mm in overall length. Crystals in the upper half of the cluster
exhibited a \textquotedblleft fishbone\textquotedblright\ morphology \cite%
{fishbones}, while slender columnar crystals grew in the lower half of the
cluster. Well developed hollow columns were seen immediately below C2.

Crystals growing between C2 and C3 (i.e. at temperatures in the range $-13$
C $<T<-8$ C) were mainly small and blocky, although some \textquotedblleft
blocky spikes\textquotedblright\ were also seen. These spikes appeared to be
equivalent to the fishbone morphology, except without distinct
sidebranching. A blocky spike thus had the appearance of a thick needle-like
crystal with stout blocky protrusions along its length.

C3 -- A well-defined cluster of planer dendritic crystals growing at $%
T\approx -15$ C. This cluster was usually visible after 10-15 minutes, with
a vertical width of 5-6 mm, implying a temperature range of $\Delta T\approx
2$ C. At one hour the longest crystals were about 10 mm in length, with tip
growth velocities near 3 $\mu $m/sec. These crystals exhibited a fern-like
dendritic structure with much sidebranching. Sectored plate crystals without
sidebranching appeared immediately above and below C3, and in low-$\sigma $
regions.

Note that the baseline ice crystal morphologies and growth rates depend on
the detailed temperature and supersaturation profiles in the diffusion
chamber. From our experience working with diffusion chambers over many
years, we have found that C2 and C3 are robust and clearly visible features
as long as the supersaturation is sufficiently high and the chamber is
clean. In other diffusion chambers with different temperature profiles,
however, C1 is not always visible as a clearly defined cluster of large
crystals.

\subsection{Chemical Additives}

Vaporous chemicals were added to the diffusion chamber by mixing with clean
nitrogen gas, such that the total flow rate through the chamber was constant
at 6.2 ml/sec, which replaced the gas in the chamber every 30 minutes. The
flow entered the top of the chamber and exited the bottom, as shown in
Figure \ref{diffchamber}. Tests with clean nitrogen gas showed that this
flow rate moved the crystal peaks down about 3 mm, reflecting a shift in air
temperature relative to the zero-flow profile, but otherwise did not
noticeably change the crystal morphologies along the string. Thus flow with
nitrogen gas only produced the baseline morphologies described above.

Two strategies were used to add vaporous chemicals to the flowing nitrogen
gas. For the lowest concentrations, a measured volume of air saturated with
chemical vapor (above the liquid phase) was injected into a 1.2 liter flask,
and a flow of 0.4 ml/sec through this flask was mixed with a nitrogen flow
of 5.8 ml/sec. The combined gas was then flowed through the diffusion
chamber. Additional saturated air was injected into the flask as needed to
keep the chemical concentration constant to within a factor of three. For
higher concentrations, a small amount of liquid was added to the flask to
saturate the air above it. Then the flow rates were adjusted to yield the
desired vapor fraction, up to the saturated vapor pressure of the chemical
being tested. Note that chemicals absorbed by the walls of the diffusion
chamber were not factored into our analysis, and \textit{in situ} chemical
concentrations were not measured. With a replacement time of 30 minutes,
however, we expect that absorption did not change the input concentrations
appreciably.

During a typical run, we first observed the baseline crystal growth in the
absence of any chemical additives to verify that the diffusion chamber was
sufficiently clean. We then increased the concentration of the chemical
additive being tested at roughly a factor of three per hour. We periodically
cleaned and replaced the string during this time, thus observing the crystal
morphologies as a function of vaporous chemical concentration in nitrogen
gas. Detailed observations are given in Appendix I for the chemical
additives tested, and conclusions from these experiments are discussed below.

\section{Flow Chamber Experiments}

\begin{figure}[h] 
  \centering
  \includegraphics[bb=0 0 990 982,width=5in,keepaspectratio]{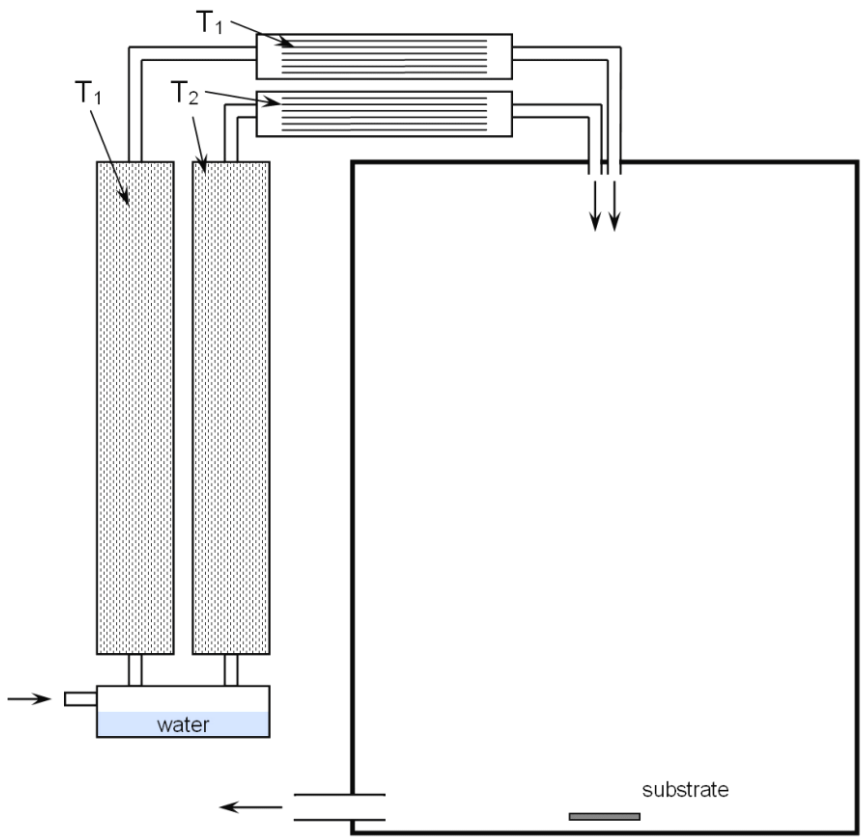}
  \caption{Schematic diagram of the flow
chamber described in the text. Gas entering the experiment is first hydrated
in a water reservoir kept near room temperature. The gas is then divided
into two flows, yielding gas saturated with water vapor at known
temperatures $T_{1}$ and $T_{2}.$ The two flows are then combined to produce
supersaturated gas at an intermediate temperature. A nucleator (not shown)
produces ice crystals that grow and eventually fall from gravity. Imaging
and broad-band interferometery are used to measure crystals that land on the
substrate \protect\cite{chamber}.}
  \label{flowchamber}
\end{figure}

In a second set of experiments we observed ice crystal growth in a free-fall
flow chamber, shown schematically in Figure \ref{flowchamber}. This
experiment was designed primarily to produce a supersaturated environment
that had the lowest achievable levels of chemical contaminants, in order to
test whether very low-level contaminants played a significant role in ice
growth. We used a high-pressure dewar of liquid nitrogen as the input gas
source, as we expected the nitrogen boil-off to be relatively uncontaminated
compared to other gas sources. (However, we did not measure the gas purity
directly in our experiments.) Warm, dry nitrogen gas from the dewar first
entered a water reservoir (see Figure \ref{flowchamber}), and tests showed
that flow over the water surface provided sufficient hydration for our
purposes. The hydrated gas then flowed through a series of heat exchangers
that yielded two gas flows that were saturated with water vapor at known
temperatures $T_{1}$ and $T_{2}$ shown in Figure \ref{flowchamber}.

The flows were then mixed in the main chamber (which was separately cooled)
to produce supersaturated gas at an intermediate temperature. A nucleator
created ice crystals that were observed as they fell onto a substrate at the
bottom of the chamber, following \cite{chamber, 2008data}. To reduce the
effects of contaminants entering the chamber via outgassing from the chamber
walls and heat exchangers, the total flow rate was kept at 1 liter/second --
sufficient to replace all the gas in the chamber every three minutes.

The supersaturation in the main chamber was found to be substantially lower
than that calculated from mixing of the two gas flows, and we attribute this
to turbulent flow causing an especially rapid interaction with the walls of
the main chamber. Since the chamber walls were coated with ice during the
experiments, this interaction acted to reduce the supersaturation from the
calculated value. Thus we used the crystal growth in air to estimate the
supersaturation, comparing our results with data from \cite{2008data}.

\subsection{Growth in Clean Nitrogen Gas}

We compared ice growth in ultra-clean nitrogen gas and ordinary air by
changing the input source while leaving everything else in the experiment
fixed. The air source was laboratory air that had been first filtered and
dehydrated, so the hydration step was the same for the two gas sources.
Although the supersaturation was not independently known in these
measurements, the similar gas paths and experimental protocols insured that
the supersaturation in nitrogen and in air were nearly identical.

\begin{figure}[h] 
  \centering
  \includegraphics[bb=0 0 868 1064,width=5in,keepaspectratio]{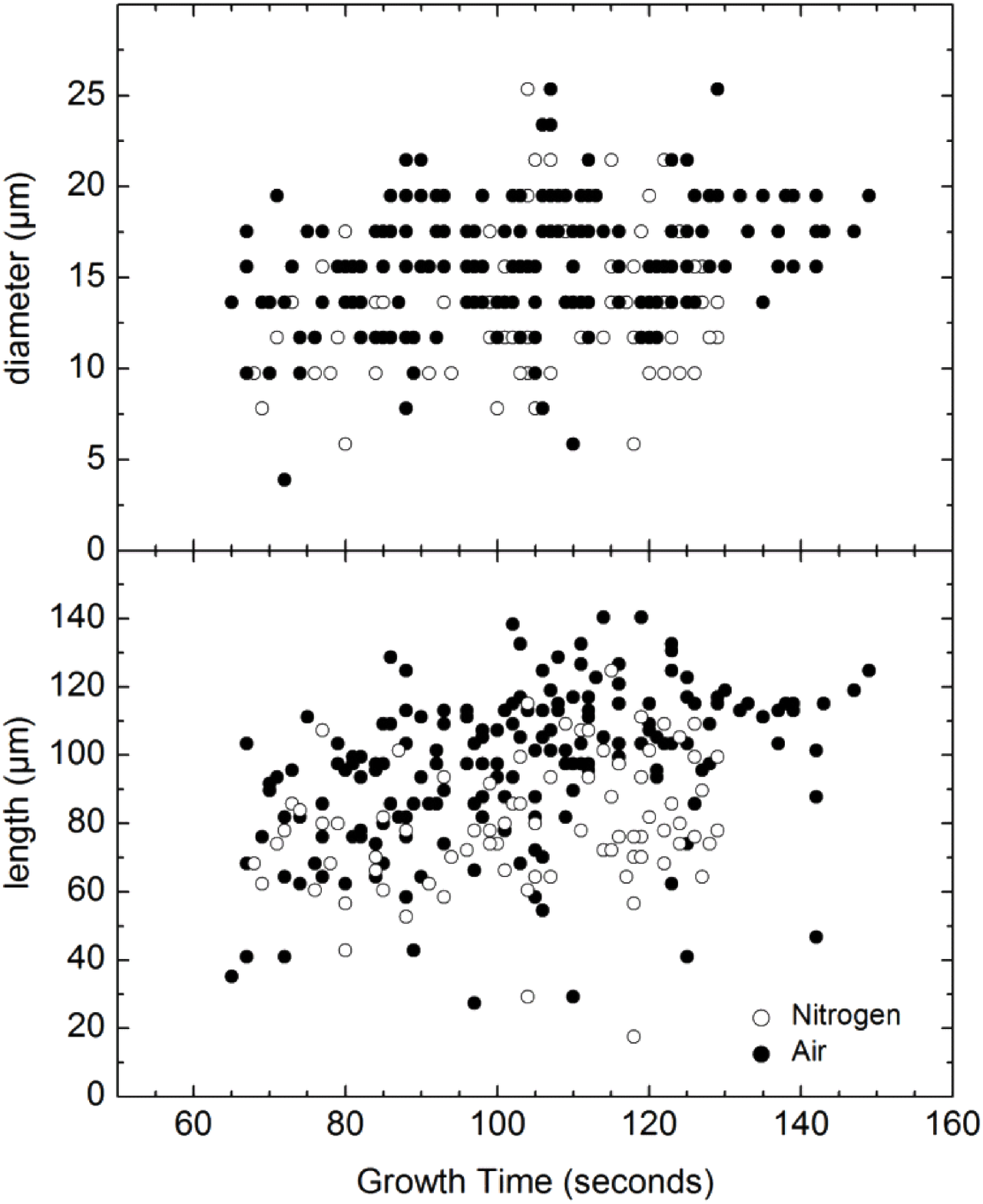}
  \caption{Diameters and lengths of ice
crystal columns grown at -5 C in the flow chamber, as a function of time
after nucleation. Solid points show growth in ordinary laboratory air, while
open points show growth in ultra-clean nitrogen gas. To the limit set by
residual systematic errors, we see no significant difference between the two
data sets.}
  \label{5Cdata}
\end{figure}

\begin{figure}[h] 
  \centering
  \includegraphics[bb=0 0 889 1096,width=5in,keepaspectratio]{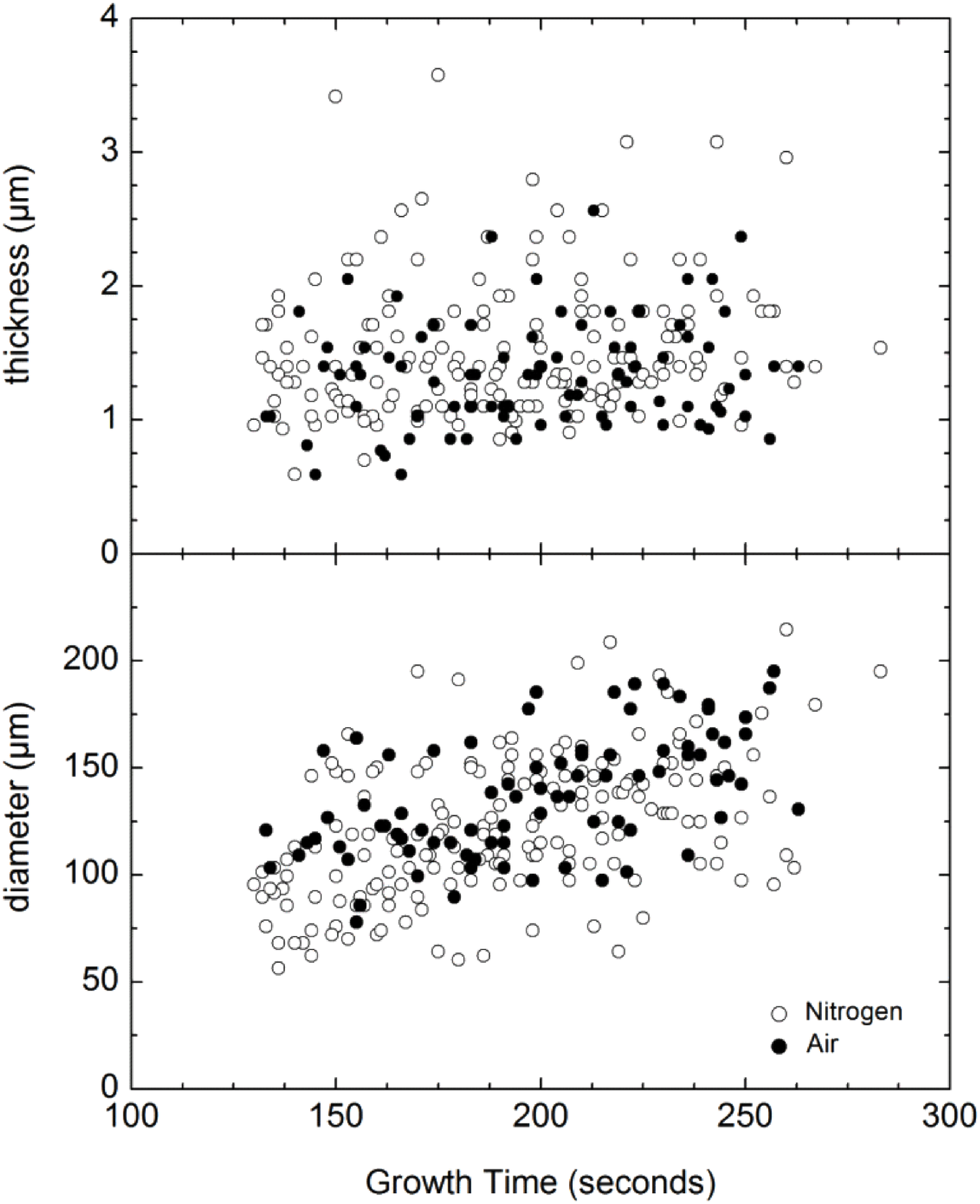}
  \caption{Thicknesses and diameters of
ice crystal plates grown at -15 C in the flow chamber, as a function of time
after nucleation. Solid points show growth in ordinary laboratory air, while
open points show growth in ultra-clean nitrogen gas. To the limit set by
residual systematic errors, we see no significant difference between the two
data sets.}
  \label{15Cdata}
\end{figure}

Data from runs at $-5$ C and $-15$ C are shown in Figures \ref{5Cdata} and %
\ref{15Cdata}, respectively. From these data we see that the growth rates
were indistinguishable to an accuracy of 10-20 percent at both temperatures.
We estimate that changes in growth at this level could be brought about by
residual systematic differences between the runs with air and with nitrogen
gas.

\subsection{Growth with Added Acetone}

We further explored the growth of plate-like crystals at -15 C in our flow
chamber as a function of added acetone vapor. A separate nitrogen gas flow
was mixed with the main flow of liquid nitrogen boil-off to produce a known
concentration of acetone in nitrogen. Although some acetone was likely
absorbed in the heat exchangers, we expect that the high flow rate produced
an acetone concentration inside the main chamber that was close to the
calculated value.

\begin{figure}[h] 
  \centering
  \includegraphics[bb=0 0 952 1182,width=5in,keepaspectratio]{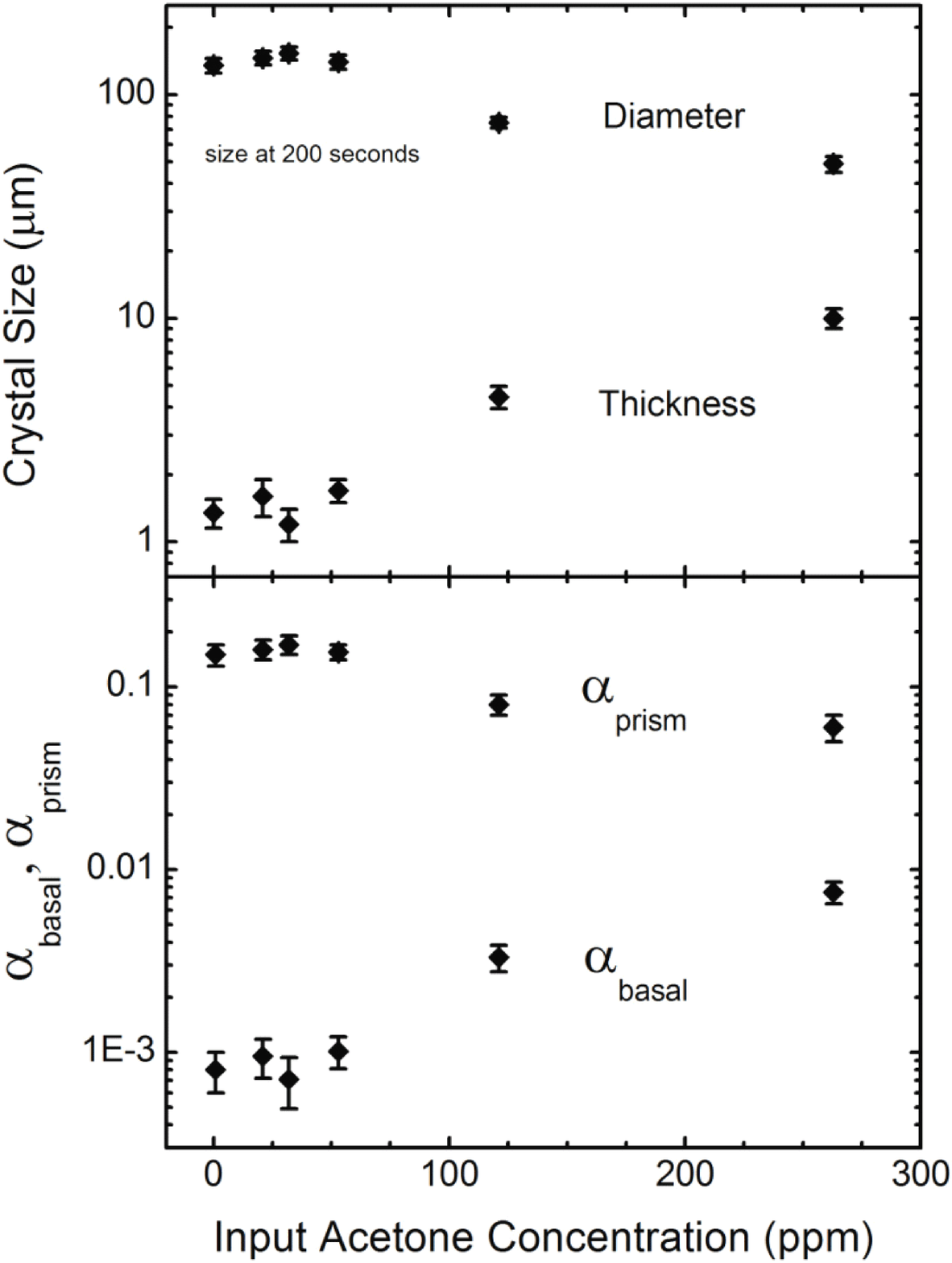}
  \caption{Data on the growth of ice crystal
plates at -15 C as a function of acetone vapor concentration in nitrogen
gas. The top panel shows average crystal diameters and thicknesses after 200
seconds of growth. The bottom panel shows condensation coefficients inferred
from the data using cellular automata modeling of diffusion-limited growth.}
  \label{acetone}
\end{figure}

Results from these measurements are shown in Figure \ref{acetone}, which
gives ice crystal diameters and thicknesses after 200 seconds of growth as a
function of acetone concentration. Note that the crystals become more blocky
with increased acetone concentration, consistent with the disappearance of
the C3 peak in the diffusion-chamber data for acetone. The flow chamber data
showed little change below 50 ppm, roughly consistent with the
diffusion-chamber data, even though the water vapor supersaturation was
substantially higher in the diffusion chamber. (The higher supersaturation
in the diffusion chamber yielded dendritic crystals with many sidebranches,
as opposed to the simple plate-like crystals seen in the flow chamber).

From the crystal sizes at low acetone concentration, along with prior data
from \cite{2008data}, we estimated that the supersaturation was
approximately 2 percent in the flow chamber data. Using this value, we then
used cellular-automata modeling of diffusion-limited growth \cite%
{libbrechtmodel} to estimate the condensation coefficients as a function of
acetone concentration, which are also shown in Figure \ref{acetone}. We see
that $\alpha _{basal}$ shows the largest change with added acetone, but we
also see that a decrease in $\alpha _{prism}$ accompanies the increase in $%
\alpha _{basal}.$

\section{Conclusions and Discussion}

From the experiments described above, we draw the following conclusions
relating to the growth of ice crystals from water vapor in the presence of
an inert background gas with added chemical vapors.

\textbf{Chemical additives below 1-10 ppm have little effect on ice growth.}
In our experiments we found that even the most active chemical additives
produced no significant changes in growth morphologies at concentrations
below 10 ppm. Other authors (see above) have reported growth changes with as
little as 1 ppm in some cases, but this may depend on supersaturation,
observation time, and other experimental differences. From the preponderance
of data, it seems reasonable to conclude that very low-level chemical
impurities, below 1-10 ppm for any chemical, likely have little effect on
ice growth under typical experimental conditions.

There are a number of caveats we must add to this conclusion. First, all
substances have not been tested, by us or others, so there may exist some
chemical vapors that affect ice growth below the 1 ppm level. Second, the
observations to date have been done at fairly high supersaturations, in
order to produce conveniently rapid growth, and we suspect that slower ice
growth might be more sensitive to the chemical environment. Third, the
observations to date have been done at fairly high temperatures, above -25
C, and we again suspect that ice growth at lower temperatures might be
affected by surface chemistry to a greater degree.

The most active chemical additives we found were the butanols -- Isobutanol
and 2-Butanol -- as both resulted in a disappearance of the C3 peak at 30
ppm. Acetone was almost as active, showing C3 disappearance at 100 ppm.

\textbf{Ice growth in very clean nitrogen gas is no different than in
ordinary air.} The timescale to form a monolayer of a strongly absorbed
chemical impurity on an ice surface can be orders of magnitude shorter than
typical crystal growth times, even at vapor concentrations of 1 ppm or lower 
\cite{impurities}. If ice growth was being changed by very low impurity
levels, however, we would have expected to see some change in growth rates
between ordinary air and clean nitrogen. The fact that no changes at all
were seen in our flow chamber data, both at -5 C and -15 C, casts serious
doubt on Libbrecht's hypothesis \cite{impurities}. Rejecting this hypothesis
leads us to the conclusion that the presence of even a chemically inert gas
at a pressure of one atmosphere greatly reduces the surface attachment
coefficients in comparison to rates at lower pressures. We suspect that an
examination of ice growth rates as a function of gas background pressure may
yield insights that help explain the morphology diagram.

\textbf{The growth of thin, plate-like dendrites at -15 C is especially
sensitive to chemical additives.} In all our diffusion-chamber data, the
disappearance of C3 was the first observable change in growth morphology as
the additive concentration was increased. The other clusters, by contrast,
were more robust and did not change their appearance appreciably until
higher concentrations were reached. In general we found that C3 was less
robust than C2, which in turn was less robust than C1.

In one run with acetone we found that the disappearance of C3 was not
proceeded by a change in the peak growth temperature. The C3 peak remained
near -15 C as dendritic growth gave way to the growth of smaller plates,
followed by the peak disappearing entirely.

This conclusion may portend a more general statement that chemical additives
have greater effects on ice growth as the temperature is reduced, although
at present we have no evidence to make this extrapolation. We did not
observe ice growth at temperatures below -25 C in our experiments. The data
do suggest, however, that it would be prudent in future experiments at very
low temperatures to consider the effects of chemical impurities in the
apparatus carefully, perhaps even at levels substantially below 1-10 ppm.

We note that in Knepp et al. \cite{knepp} the authors did not observe the C3
peak in their diffusion chamber during clean air runs. The fact that C3 is
particularly sensitive to impurities suggests possible contamination
problems in the Knepp et al. experiments.

\textbf{Chemical additives usually promote columnar morphologies over
plate-like morphologies.} In our diffusion-chamber observations (see the
Appendix), we often observed blocky spikes and/or columns at temperatures
near -15 C, where plate-like crystals are normally observed. In some cases,
especially at lower temperatures, the chemically induced columns were long
and thin enough to be called needles. In contrast, we never saw a set of
circumstances for which chemical additives promoted the growth of large thin
plates at any temperature. The largest, thinnest plate-like crystals
appeared only near -15 C and only in inert gases with no chemical additives.

\textbf{Chemical activity for affecting ice growth has little correlation
with solubility in water.} The two chemicals we tested with the lowest
solubilities in water -- Dichloromethane and Xylenes -- also produced the
smallest changes in ice growth (i.e., these chemicals required the highest
vapor concentrations before C3 disappeared). On the other hand, the butanols
have fairly low solubilities yet were the most active chemicals tested. From
the relatively small number of chemical vapors tested, it appears that
solubility in water is not a dominant factor in determining chemical
activity for affecting ice growth.

\textbf{Many inert gases yield similar ice growth morphologies.} In our
experiments with air, nitrogen, argon, helium, hydrogen, and methane we
observed essentially identical ice growth characteristics. Carbon dioxide
did affect ice growth to a small degree, as described in the Appendix.

In summary, we have observed the effects of vaporous chemicals on ice
growth, yielding the set of conclusions above. Our main motivation for this
study was to obtain a more quantitative knowledge relating to how chemical
impurities can affect ice growth experiments, in order to reduce unforeseen
systematic errors in future measurements. We believe that our observations
will be quite useful in this regard, as a guide to designing better
experiments investigating the molecular dynamics of ice crystal growth.

Although chemical additives exhibit some essentially universal behavior in
their effects on ice growth, our theoretical understanding of the underlying
mechanisms responsible for these growth changes is quite poor. Additional
experiments measuring ice growth rates as a function of temperature,
supersaturation, and chemical concentration may yield additional interesting
behaviors and insights beyond those found in the present experiments.

\section{Appendix I - Results from Diffusion Chamber Observations}

What follows are descriptions of ice crystal morphologies in comparison to
the baseline morphologies (described above) seen in our diffusion chamber
after one hour of growth. Morphologies are the same as in the baseline
unless otherwise noted. For each chemical additive, the concentration in
nitrogen gas was increased slowly with time while the string was
periodically cleaned and replaced. Concentration refers to the partial
pressure of the additive vapor divided by one bar.

\subsection{Acetic Acid}

Vapor pressure at 25 C = 16 Torr; miscible in water

\noindent100 ppm or less -- No observable change from normal growth

\noindent300 ppm -- C3 dendrite cluster gone, replaced by sparse blocky
spikes

\qquad normal growth for $-7$ C $<T<-3$ C

\qquad C1 appeared more quickly than normal, with enhanced sidebranching

\noindent1000 ppm -- same as 300 ppm, but C1 especially fernlike

\noindent3000 ppm -- all fernlike dendrites (C1) for $T>-12$ C

\qquad fishbone variants and blocky spikes at lower temperatures

\noindent10000 ppm -- liquid drops for $T>-4$ C

\qquad small C1 dendrites without prism facets near $T\approx -6$ C

\qquad fernlike C1 dendrites for $-10$ C $<T<-6$ C

\qquad fishbone variants for $T<-10$ C, with no retrograde sidebranches

\subsection{Acetone}

Vapor pressure at 25 C = 240 Torr; miscible in water

\noindent10 ppm or less -- No observable change from normal growth

\noindent30 ppm -- C3 growth slower, with plate-like sidebranching

\noindent100 ppm -- C1, C2 normal

\qquad C3 gone, replace with small plates for $T<-12$ C, larger plates for $%
-12$ C $<T<-5$ C

\noindent300 ppm -- C1 growth slower than normal

\qquad C2 normal; some stout columns between C1 and C2

\qquad stout hollow columns below C2; thinner hollow columns for $T<-15$ C

\noindent1000 ppm -- roughly same as at 300 ppm

\noindent3000 ppm -- same

\noindent10000 ppm -- C1, C2 about same as at 300 ppm

\qquad single broad cluster for $-20$ C $<T<-6$ C, including fishbones with
thin, blade-like sidebranching

\noindent 30000 ppm -- liquid droplets nucleate at all temperatures

\subsection{Dichloromethane}

Vapor pressure at 25 C = 400 Torr; solubility 1.3\%

\noindent30000 ppm or less -- No observable change from normal growth

\noindent10 percent -- C1 and C2 normal; C3 gone

\qquad stout hollow columns for $-18$ C $<T<-7$ C

\qquad thin hollow columns for $T<-18$ C

\noindent50 percent -- blocky plates for $-5$ C $<T<-1$ C

\qquad needle clusters and fishbones for $-9$ C $<T<-5$ C

\qquad stout hollow columns for $-16$ C $<T<-11$ C

\qquad thin hollow columns and needles for $T<-19$ C

\subsection{Ethanol}

Vapor pressure at 25 C = 60 Torr; miscible in water

\noindent300 ppm or less -- No observable change from normal growth

\noindent1000 ppm -- C3 replaced by a broad cluster of sectored plates

\qquad C2 appears at $T\approx -3.5$ C, with normal appearance

\noindent3000 ppm -- same as 1000 ppm, but C3 gone, replaced by blocky
spikes and small plates

\noindent10000 ppm -- liquid droplets nucleate on string at all temperatures

\qquad on already nucleated ice, C1 dendrites show enhanced branching,
reduced prism faceting

\subsection{Isobutanol}

Vapor pressure at 25 C = 10 Torr; solubility 9\%

\noindent10 ppm or less -- No observable change from normal growth

\noindent30 ppm -- C1, C2 normal

\qquad C3 initially showed sectored-plate sidebranches; later gone, replaced
by small blocky spikes and small sectored plates

\noindent100 ppm -- C1, C2 clusters normal

\qquad stout hollow columns for $-20$ C $<T<-5$ C

\qquad thin hollow columns and needles for $T<-20$ C

\noindent300 ppm -- C1 normal

\qquad C2 replaced by thin blocky spikes

\qquad columns and needles for $T<-5$ C, as at 100 ppm

\noindent1000 ppm -- small planer dendrites with no prism faceting for $-2$
C $<T<0$ C

\qquad otherwise same as at 300 ppm

\noindent3000 ppm -- small planer dendrites with no prism faceting for $-2$
C $<T<0$ C

\qquad C1 and C2 merged into broad cluster of spikes

\qquad stout hollow columns for $T<-7$ C (no longer thin needles at lower
temperatures)

\noindent10000 ppm -- small leafy planer dendrites for $-4$ C $<T<-1$ C;
droplets above

\qquad small spikes and sectored plates at lower temperatures

\subsection{2-Butanol}

Vapor pressure at 25 C = 13 Torr; solubility 29\%

\noindent10 ppm or less -- No observable change from normal growth

\noindent30 ppm -- C1, C2 normal

\qquad C3 gone, replaced by shorter dendrites and plates

\noindent100 ppm -- C1, C2 clusters normal

\qquad stout hollow columns for $-15$ C $<T<-5$ C

\qquad thin hollow columns and needles for $T<-15$ C

\noindent300 ppm -- C1 normal

\qquad C2 crystals shorter than normal, with branching at smaller angular
separation

\qquad same as 100 ppm for $T<-5$ C

\noindent1000 ppm -- C1 and C2 merged into single cluster of spikes for $-7$
C $<T<-1$ C

\qquad shorter spikes with end-plates close to $T=0$

\qquad same as 100 ppm for $T<-5$ C

\noindent3000 ppm -- small planer dendrites with no prism faceting for $-2$
C $<T<0$ C

\qquad C1 and C2 merged into broad cluster of spikes at $T\approx -4$ C

\qquad small plates mixed with spikes and blocky crystals for $T<-5$ C

\noindent10000 ppm -- well developed planer dendrites for $-4$ C $<T<-1$ C

\qquad most large spikes and smaller blocky crystals for $-12$ C $<T<-5$ C

\qquad small sectored plates for $T<-15$ C

\subsection{Isopropanol}

Vapor pressure at 25 C = 44 Torr; miscible in water

\noindent30 ppm or less -- No observable change from normal growth

\noindent100 ppm -- C3 shows smaller, fewer, more plated dendrites, with
large plates between dendrites

\noindent300 ppm -- C3 gone, replaced with spikes and small plates

\qquad C1 and C2 normal; thin hollow columns between C1 and C2

\qquad stout hollow columns and plates for $-9$ C $<T<-5$ C

\noindent1000 ppm -- As above for $T>-9$ C

\qquad spikes and some small plates for $-16$ C $<T<-9$ C

\qquad then hollow columns for $T<-16$ C

\noindent3000 ppm -- C1 and C2 gone

\qquad stout hollow columns for $-12$ C $<T<-3$ C

\qquad needles and hollow columns for $-18$ C $<T<-12$ C

\qquad narrow hollow columns for $T<-18$ C

\noindent10000 ppm -- spikes with fine dendrites for $-5$ C $<T<-2$ C

\qquad blocky spikes and small plates for $-14$ C $<T<-5$ C

\qquad small plates for $T<-14$ C

\noindent30000 ppm -- only droplets observed on string at all temperatures

\subsection{Xylenes}

Vapor pressure at 25 C = 9 Torr; solubility 0.02\%

\noindent1000 ppm or less -- No observable change from normal growth

\noindent3000 ppm -- C3 weak; dendrites appear with sectored-plate branches

\noindent10000 ppm -- C3 gone, replaced by blocky spikes

\qquad C1, C2 normal

\subsection{Pure Gases}

In several runs with air, nitrogen, argon, helium, hydrogen, and methane
gases at essentially 100 percent concentration in the diffusion chamber we
did not observe substantial changes in ice growth morphologies. With carbon
dioxide in the chamber we found that C1 and C2 were normal, while C3 was
gone, being replaced by spikes and other small crystals.

\end{document}